\documentclass[prl,reprint,showpacs,amsmath,amssymb]{revtex4-1}
\usepackage{graphicx,color,amsmath,amssymb} 
\usepackage[colorlinks=true,urlcolor=blue, linkcolor=blue]{hyperref}

\def\({\left(}
\def\){\right)}

\def\ad{a^\dagger}

\def\eqs#1#2{Eqs.~(\ref{eq:#1}-\ref{eq:#2})} 
 
\def\fig#1{Fig.\ref{fig:#1}}
\def\figs#1#2{Figs.\ref{fig:#1} \& \ref{fig:#2}}
\def\Fig#1{Figure~\ref{fig:#1}}

\begin{document}

\title{Parallel generation of quadripartite cluster entanglement in the optical frequency comb}
\author{Matthew Pysher,$^1$ Yoshichika Miwa,$^{2}$
 Reihaneh Shahrokhshahi,$^{1}$ Russell Bloomer,$^{1}$ and Olivier Pfister$^{1 \ast}$\\
{\em $^1$Department of Physics, University of Virginia, Charlottesville, Virginia 22903, USA\\
$^2$Department of Applied Physics, School of Engineering, The University of Tokyo, 7-3-1 Hongo, Bunkyo-ku, Tokyo 113-8656, Japan}}


\date{June 20, 2011}

\begin{abstract}
Scalability and coherence are two essential requirements for the experimental implementation of quantum information and quantum computing. Here, we report a breakthrough toward scalability: the simultaneous generation of a record 15 quadripartite entangled cluster states over 60 consecutive cavity modes (Qmodes), in the optical frequency comb of a single optical parametric oscillator. The amount of observed entanglement was constant over the 60 Qmodes, thereby proving the intrnisic scalability of this system. The number of observable Qmodes was restricted by technical limitations, and we conservatively estimate the actual number of similar clusters to be at least three times larger. This result paves the way to the realization of large entangled states for scalable quantum information and quantum computing.
\end{abstract}

\maketitle

\paragraph*{Introduction}

The experimental implementation of quantum computing, driven by the promise of exponential speedup for tasks such as the  simulation of quantum physics \cite{Feynman1982} and integer factoring \cite{Shor1994} is a daunting challenge that requires exquisite levels of control over the quantum mechanical properties of numerous individual physical systems (quantum bits or, in this paper, quantum modes or Qmodes). The response to this challenge spawned a wealth of experimental research efforts in widely different fields \cite{Ladd2010}, striving to enable and maintain quantum-coherent temporal evolution of quantum bits, while at the same time scaling up their number. Here, we demonstrate a breakthrough toward scalability: the novel, ultracompact implementation of quantum registers in the optical frequency comb (OFC) formed by the spectrum of a {\em single} optical parametric oscillator (OPO), thereby utilizing a capability for quantum information storage analogous to that exploited classically in FM radio or wavelength multiplexing. The classical OFC  generated by ultrastable pulsed lasers has found groundbreaking uses in ultimate precision frequency measurements \cite{Hansch2006,Hall2006}. In the case of the quantum OFC, each (Q)mode  is well approximated by a quantum harmonic oscillator whose continuous-variable Hilbert space is defined by its amplitude- or phase-quadrature field observable (analogues of the position and momentum observables). There is no known fundamental impossibility to the implementation of quantum computing with Qmodes \cite{Lloyd1999,Bartlett2002,Menicucci2006}, even though the implementation of quantum error correction appears likely to require Hilbert-space discretization \cite{Niset2009,Ohliger2010}. A method to create a  frequency-degenerate $N$-Qmode register was proposed, by use of $N$ OPOs and a $2N$-port interferometer \cite{vanLoock2000}, and demonstrated \cite{Aoki2003,Jing2003} for 3 and 4 Qmodes. However, it was also shown that a square-grid continuous-variable cluster state of arbitrary size, suitable  for universal one-way quantum computing \cite{Raussendorf2001,Menicucci2006}, can be generated in the OFC of a single OPO \cite{Menicucci2008,Flammia2009}. In this work, we achieved the first step toward this goal: the parallel generation of 15 quantum computing registers, each comprising 4 Qmodes in a quadripartite cluster state, in the quantum OFC of a single OPO. Requirements for the generation of larger entangled states include the experimental progress made in this work, along with a richer pump spectrum and a more tailored nonlinear interaction \cite{Menicucci2008,Flammia2009}.

\paragraph*{Experimental method} 

The quantum OFC was generated by a bowtie ring OPO containing two $x$-cut $\rm KTiOPO_4$ (KTP) nonlinear crystals, of 10 mm length, and rotated by $90^\circ$ from each other about the $x$ axis. This ensured the perfect overlap of the respective OFCs of orthogonal linear polarizations $y$ and $z$. One crystal was not phasematched. The other was periodically poled with two distinct periods: 9 $\mu$m over a 3 mm length, and 458 $\mu$m over 7 mm. The former quasiphasematched the $zzz$ parametric downconversion, where the first letter denotes the polarization of the pump field at frequency $2\omega_o$ and the other letters denote the polarization of the $n^\mathrm{th}$ signal field pair at $\omega_{\pm n}$=$\omega_o$$\pm$$(n-1/2)\Delta$, with $\Delta$=945.66 MHz the free spectral (FSR) range of the OPO cavity. The latter period quasiphasematched the $yzy$ and $yyz$ interactions simultaneously (dispersion was negligible for our values of $n$). The pump polarization was carefully adjusted in the $(yz)$ plane, using OPO characterization by resonant second harmonic generation \cite{SuppInfo}, to yield the Hamiltonian \cite{Zaidi2008}
\begin{equation}
H=i\hbar\kappa \sum_n \(\ad_{-n,z}\,\ad_{n,z}+ \ad_{-n,y}\, \ad_{n,z}+ \ad_{-n,z}\, \ad_{n,y}\)+H.c.
\label{eq:ham}
\end{equation}
where $\ad_{j,k}$ is the creation operator of the $k$-polarized Qmode of frequency $\omega_{j}$. This Hamiltonian entangles the OFC as depicted in \fig{squares}
\begin{figure}
\centerline{\includegraphics[width=1.8in]{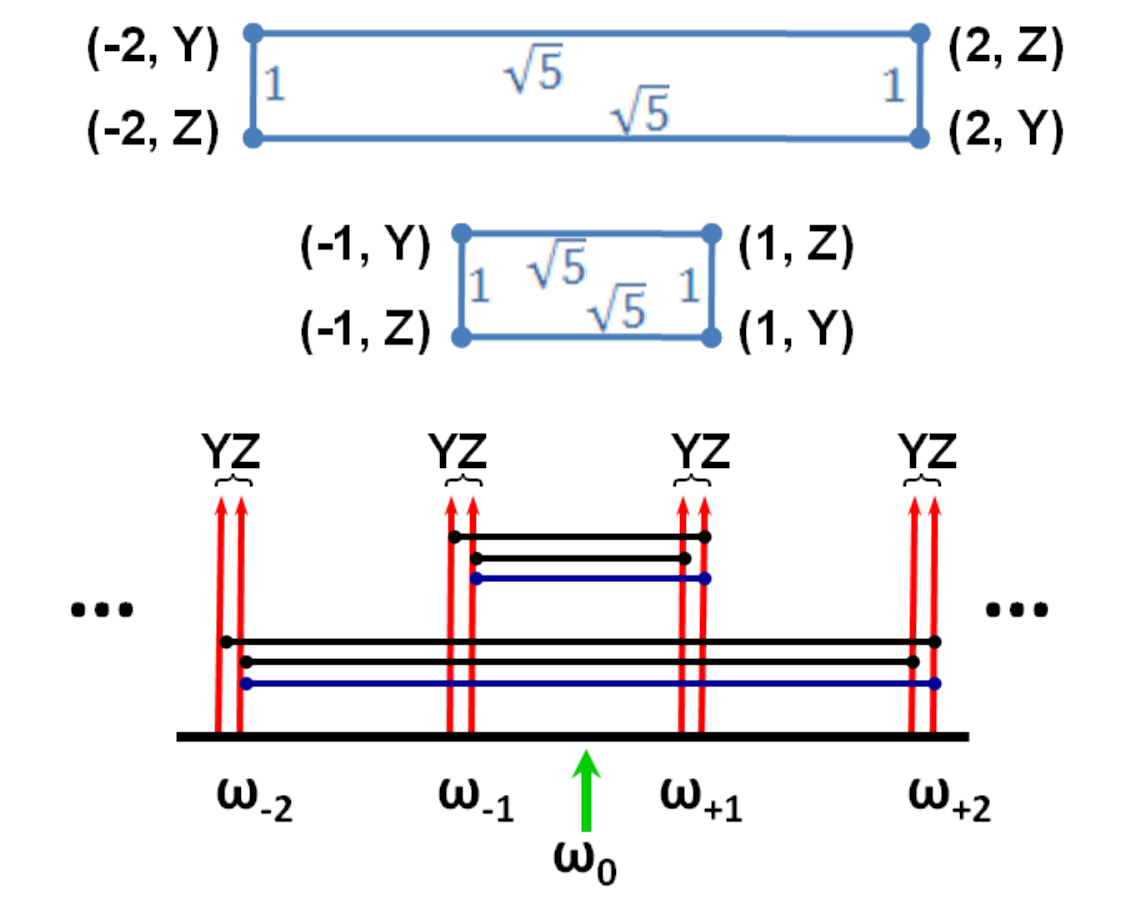}}
\caption{Principle of the experiment. The OFC of a single OPO is made polarization-degenerate by using a cavity with 2 identical crystals rotated 90 degrees from each other in the polarization plane. One crystal simultaneously phasematches the $zzz$, $yzy$, and $yyz$ nonlinear interactions, bottom. This creates square weighted cluster states, top, in blue \cite{Zaidi2008}.}
\label{fig:squares}
\vskip -0.2in
\end{figure} 
and proven by the solutions of the Heisenberg equations for the $n^\mathrm{th}$ Qmode quartet:  
\begin{align}
\label{eq:q+}
Q_+ & = \left\{\left[Q_{-n,y} - Q_{n,y}\right] + \Phi\left[Q_{-n,z} - Q_{n,z}\right]\right\} e^{-r\Phi}\\
\label{eq:p+}
P_+ & = \left\{\left[P_{-n,y} + P_{n,y}\right] + \Phi\left[P_{-n,z} + P_{n,z}\right]\right\} e^{-r\Phi}\\
\label{eq:q-}
Q_- & = \left\{\Phi\left[Q_{-n,y} + Q_{n,y}\right] - \left[Q_{-n,z} + Q_{n,z}\right]\right\} e^{-\frac r\Phi}\\
\label{eq:p-}
P_- & = \left\{\Phi\left[P_{-n,y} - P_{n,y}\right] - \left[P_{-n,z} - P_{n,z}\right]\right\} e^{-\frac r\Phi}
\end{align}
where $Q=a+a^\dag$ and $P=i(a^\dag-a)$ are the amplitude and phase quadratures, $r$ is the squeezing parameter, and $\Phi=(\sqrt 5 +1)/2$, which is the golden ratio. These  four squeezed (quantum-noise reduced) field quadratures coincide, to local quadrature phase shifts left, with the nullifiers (entanglement witnesses) of a weighted square cluster state \cite{Zaidi2008} (\fig{squares}) in the (unphysical) limit of infinite squeezing, where the cluster state is a zero-eigenvalue eigenstate of the nullifiers. The exponentiated nullifiers are thus the stabilizers of the entangled state \cite{Gu2009} (in contrast to the Qbit case, weighted Qmode cluster states are stabilizer states \cite{Menicucci2011}). For a pure state, observing the squeezing of a nullifier suffices to prove that the state has been prepared into a stabilizer state. For a statistical mixture, the situation is more complicated but one can still use the van Loock-Furusawa criteria \cite{vanLoock2003a} to prove quadripartite nonseparability. We experimentally demonstrated both.

The setup is described in \fig{setup}. The OPO, pumped at 532 nm by a frequency-doubled, ultrastable continuous-wave Nd:YAG laser (Innolight Diabolo), consisted in a cavity with low-loss mirrors and a 5\% output coupler. The quantum OFC was separated into its orthogonal polarizations, and quadrature combinations, e.g.\ $Q_{-n,y}$$\pm$$Q_{n,y}$ in \eqs{q+}{p-}, were measured by two-tone balanced homodyne detection with local oscillator (LO) fields at $\omega_{\pm n}$. 
\begin{figure}
\centerline{\includegraphics[width=3.4in]{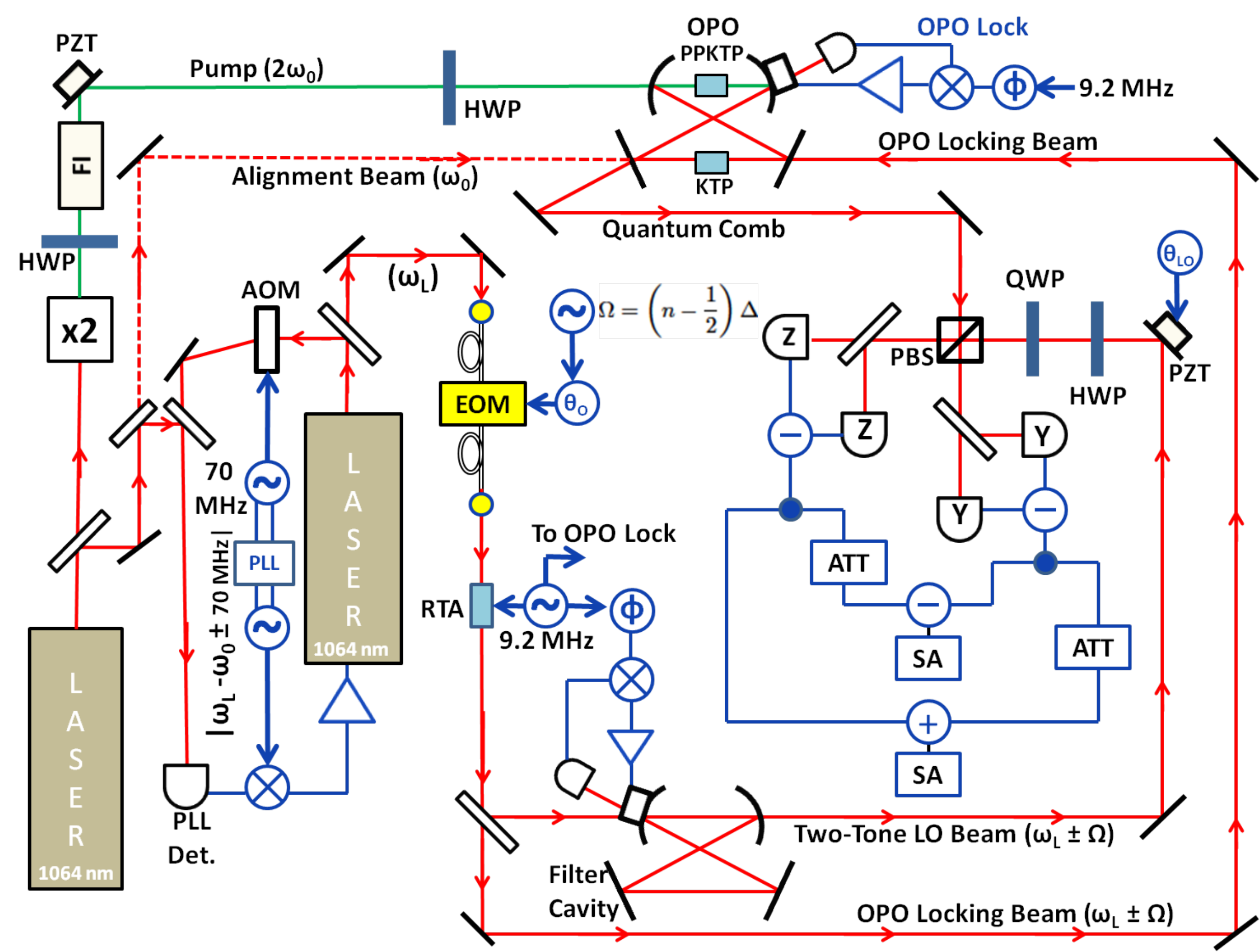}}
\caption{Experimental setup with HWP, half waveplate; QWP, quarter waveplate; FI, Faraday isolator; PZT, piezo-electric transducer; PLL, phase-lock loop; AOM, acousto-optic modulator; EOM, waveguide electro-optic modulator; KTP, $\rm KTiOPO_4$; RTA, $\rm RbTiAsO_4$ EOM; PBS, polarizing beam splitter; ATT, RF attenuator; SA, spectrum analyzer.}
\label{fig:setup}
\vskip -0.2in
\end{figure} 
The LO originated from another Nd:YAG laser (Lightwave Electronics) which was phaselocked to the 1064 nm pump laser before it was frequency doubled \cite{SuppInfo}. The LO laser frequency could therefore coincide with $\omega_o$ (\fig{data}), or differ from it for experimental verifications (\fig{check}). The $\omega_{\pm n}$ frequencies were then generated  by  phase electro-optic modulation (EOM) and subsequent bandpass filtering by an optical cavity of same FSR as the OPO, in order to remove the carrier and second harmonics. The homodyne visibilities were  97\% for the $y$ polarization and 96\% for $z$. Finally, the homodyne photocurrents from 95\% efficient InGaAs photodiodes (JDSU ETX500T) were preamplified and combined by RF splitters and attenuators in order to yield the variance of nullifiers (\ref{eq:q+},\ref{eq:p-}), observed synchronously on two spectrum analyzers while the LO optical path $\theta_{LO}$ was scanned. The measured observables can be expressed in terms of generalized quadratures $A(\theta)=ae^{-i\theta}+\ad e^{i\theta}$ as 
\begin{align}
A_\pm(\theta) =&\left\{\left[A_{-n,y}(\theta) \mp A_{n,y}(-\theta)\right] \right.\nonumber\\
& \pm \left.\Phi^{\pm 1}\left[A_{-n,z}(\theta) \mp A_{n,z}(-\theta)\right]\right\} e^{-r\Phi^{\pm 1}}
\end{align}
where phase values $\theta$=0,$\pi/2$ yield amplitude and phase quadratures, respectively. Note that the squeezing is independent of $\theta$. Because  we use two-tone homodyne detection, $\theta$ is a function of $\theta_{LO}$ and of the  EOM phase $\theta_o$ (\fig{setup}), and the different nullifiers \eqs{q+}{p-} are obtained for the respective values ($\theta_{LO}$,$\theta_o$)=(0,0);(0,$\pi/2$);($\pi/2$,$\pi/2$);($\pi/2$,0), modulo $\pi$. Additional checks were made using LO polarization \cite{SuppInfo}.

\paragraph*{Results}
Fifteen sets of 4 Qmodes were measured for $n$=1 to 15. The measurement results are displayed in \fig{data}. 
\begin{figure}
\centerline{\includegraphics[width=3.4in]{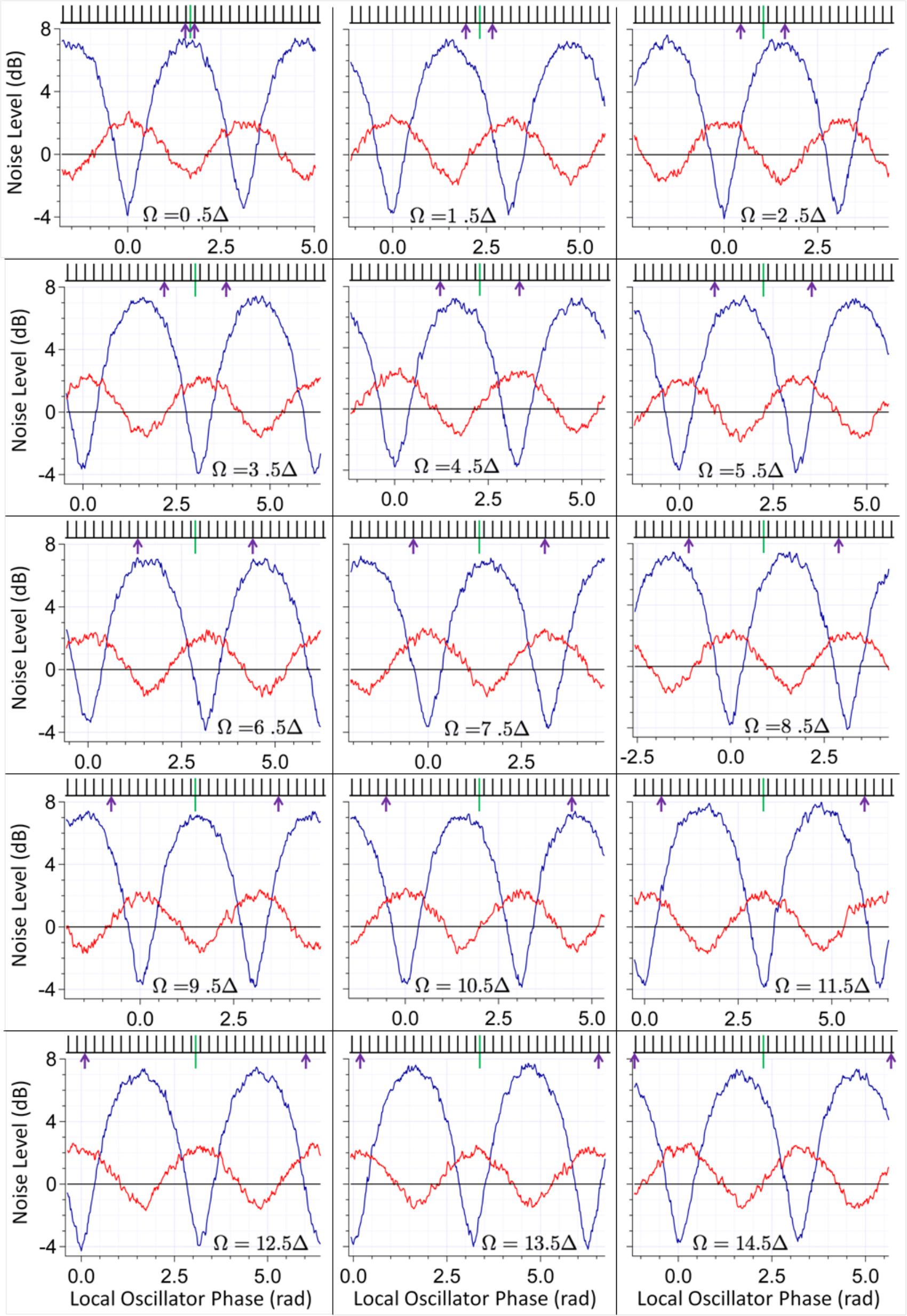}} 
\caption{Scaling quadripartite entanglement in the optical frequency comb of a single OPO: nullifier variance $\Delta A_\pm[\theta(\theta_{LO})]^2$, relative to the vacuum noise, versus the LO phase $\theta_{LO}$. Note that the squeezing depends on $\theta_{LO}$ but not on $\theta$ \cite{SuppInfo}. Single-sweep measurements were taken at 1.25 MHz frequency; RBW: 30 kHz; VBW: 30 Hz. The Qmodes (black lines) measured are marked by the LO sidebands (purple arrows). The green line references half the pump frequency.}
\label{fig:data}
\vskip -0.2 in
\end{figure} 
As can clearly be seen, the level of squeezing is constant over the whole set of 15 observed clusters, which establishes scalability in the OFC. Moreover, the maximum value of $n=15$ was not fixed by the quantum state preparation process but by measurement limitations: the 14 GHz bandwidth limit of the EOM. The state preparation bandwidth is given by the phasematching bandwidths of the nonlinear interactions. We calculated these for the $zzz$ and $yzy$ processes, respectively, to be 616 and 47 GHz at 99\% of the maximum. This indicates that a 1\% squeezing decrease occurs for the $yzy$ interaction at 47 GHz (a 10\% decrease at 153 GHz). This is too weak an effect to be observed at our current squeezing level and cluster states $n=16,...,47$ should therefore be identical to the ones measured on \fig{data}. We therefore expect that three to ten times as many cluster states were generated than the 15 that were accessible with our setup. We can also flat-top shape the phasematching curve \cite{Fejer1992} in order to optimize scalability. 

Phase-locking our two 1064 nm lasers to each other allowed us to make the crucial checks necessary to test the validity of our experimental results, in particular of our two-tone homodyne detection. These checks consisted of: using a single LO sideband, placing the LO sidebands at uncorrelated frequencies, and detuning the pump frequency from our quantum OFC. \Fig{check} shows typical results, which all agree with theoretical predictions \cite{SuppInfo} and clearly show no quantum correlations whatsoever, in stark contrast with the nullifier squeezing signals of \fig{data}.
\begin{figure}
\centerline{\includegraphics[width=3.4in]{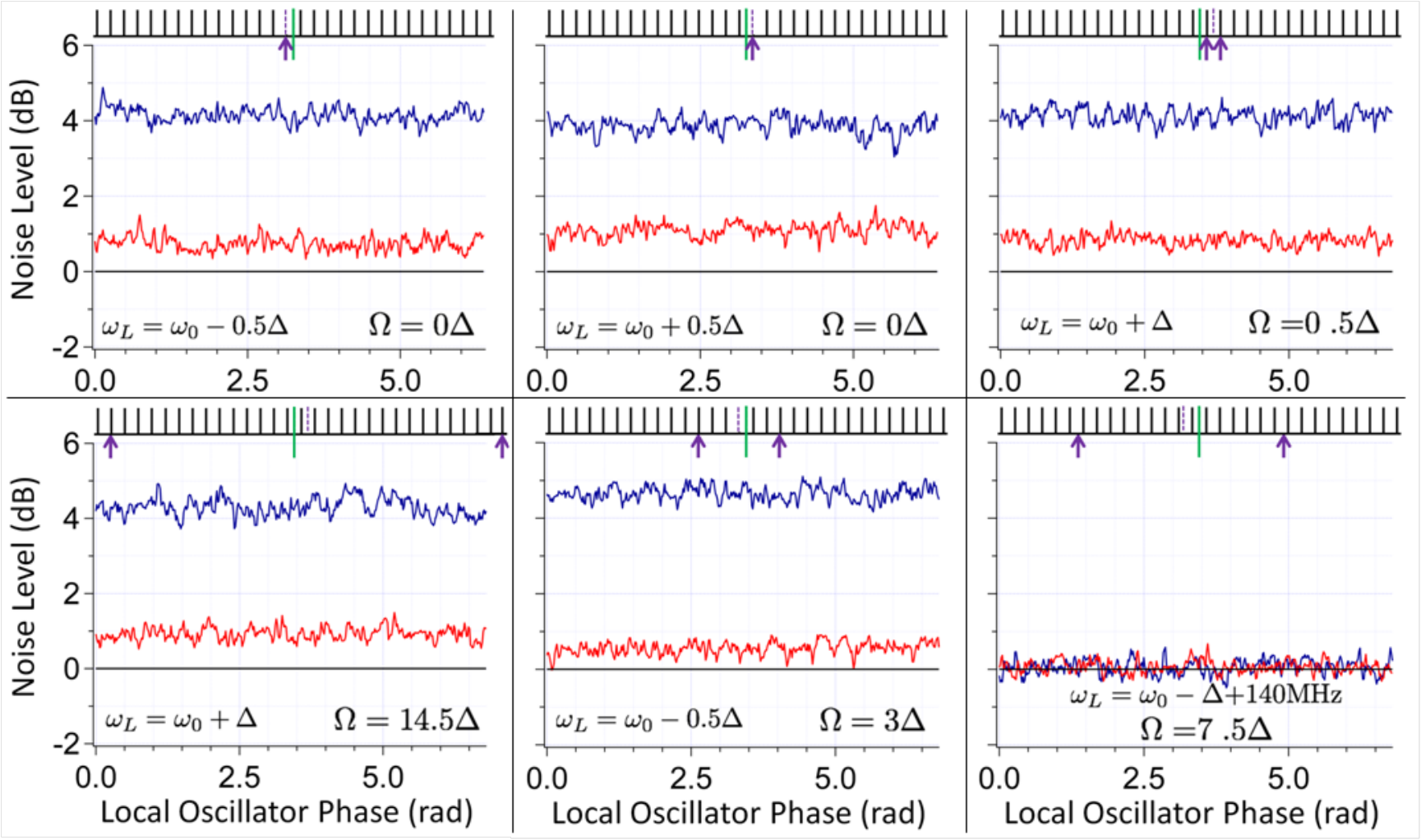}}
\caption{Detection and entanglement checks. As in \fig{data}, the plots display nullifier variance measurement relative to the vacuum noise versus the LO phase $\theta_{LO}$. Top row, left and center, single-sideband detection displays no single-mode squeezing in the OPO comb, in any quadrature.Top row, left, and bottom row, left and center, LO sidebands coincide with uncorrelated comb lines, which yields no multimode squeezing whatsoever, again no matter the LO phase used. Bottom row, right, OPO pump detuned from the comb ($2\omega_o\neq\omega_{-n}+\omega_n$) which makes the nonlinear interaction singly (instead of doubly) resonant and yields negligible squeezing. Notations and legends as in \fig{data}. The dashed purple line references the phase-locked laser's frequency.}
\label{fig:check}
\vskip -0.2 in
\end{figure} 
An essential point here is that all these verification results were insensitive to the LO phase, unlike the nullifier measurements.

We finally address pure state preparation. The fact that the antisqueezing magnitude is larger than the squeezing one points to the existence of losses and additional classical noise (from the pump laser), and therefore to the creation of a statistical mixture rather than a pure state. This can be alleviated by filtering the pump field with a ``mode-cleaner'' cavity, which we didn't do so as to maximize the pump power and hence the amount of squeezing. Nullifier squeezing is enough to claim entanglement if the state is pure.  In order to ascertain this, we measured the squeezing spectra of $A_\pm$ at the optimum phases, \fig{spectra}.
\begin{figure}
\begin{center}
\includegraphics[width=3.4in]{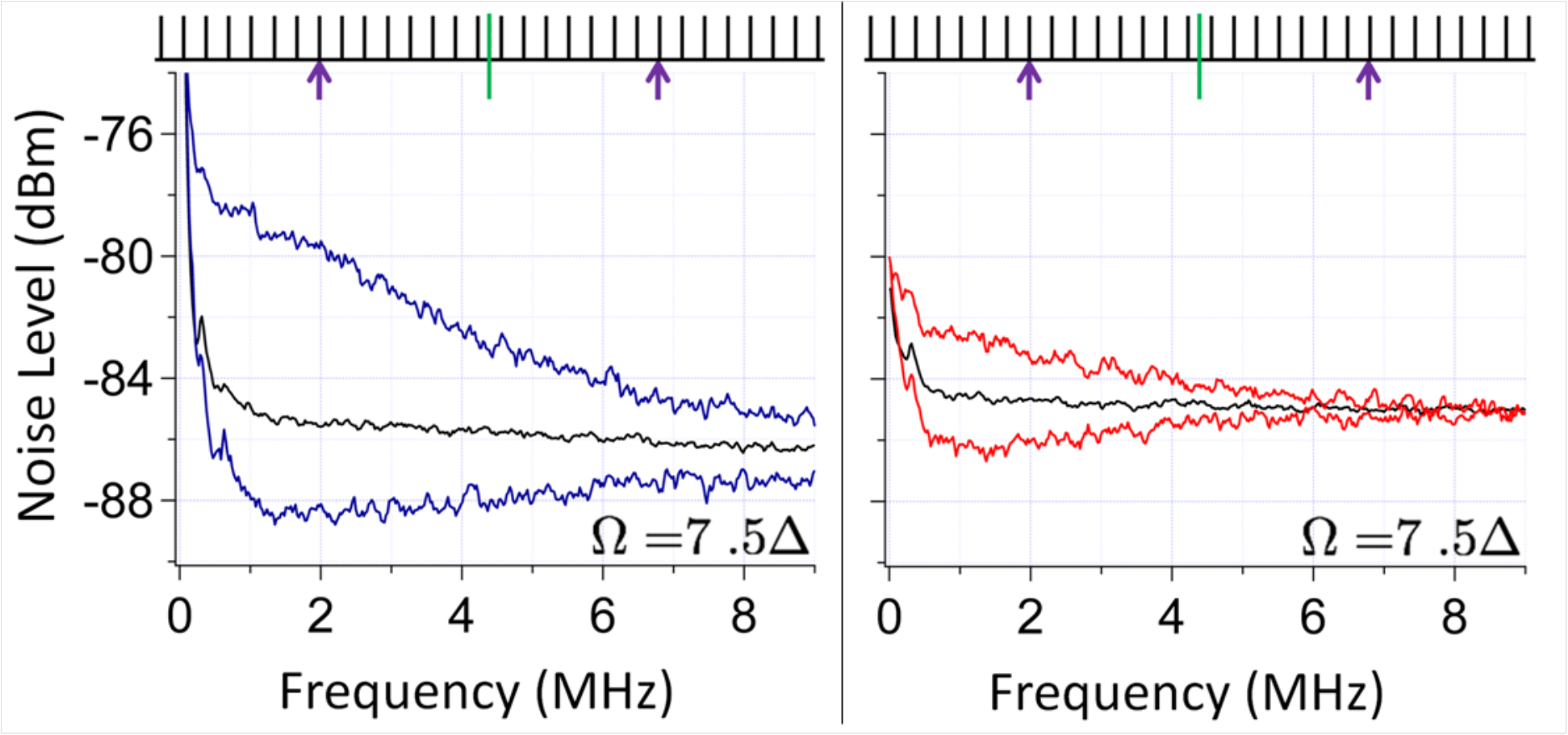}
\caption{Squeezing spectra for $A_+$, left, and $A_-$, right. The squeezed trace on the left was recorded simultaneously with the antisqueezed trace on the right, for $\theta_{LO}=\pi/2$, and vice versa, for $\theta_{LO}=0$, as with measurements in \figs{data}{check}.}
\label{fig:spectra}
\end{center}
\vskip -0.2 in
\end{figure}
As can be seen, the state is pure for measurement frequencies above 5 MHz \cite{SuppInfo}, which validates our cluster-state preparation claim. 

In the case of a mixture, as in the case of our 1.25 MHz measurement frequency (which yields more squeezing), one can use the van Loock-Furusawa (vLF) separability criterion \cite{vanLoock2003a} in order to show that no Qmode can be placed in a  factorized density operator of its own. A detailed analysis \cite{SuppInfo} led to five vLF inequalities that must all be  experimentally violated. While some of these inequalities have bounds at the vacuum level and are trivially violated by mere nonzero squeezing, others have bounds below the vacuum noise and therefore present higher violation thresholds. \Fig{vLF} displays experimental results for the two most difficult such cases, which were clearly violated, thereby  proving quadripartite entanglement even in a mixed state. 
\begin{figure}[htbp]
\begin{center}
\includegraphics[width=2.in]{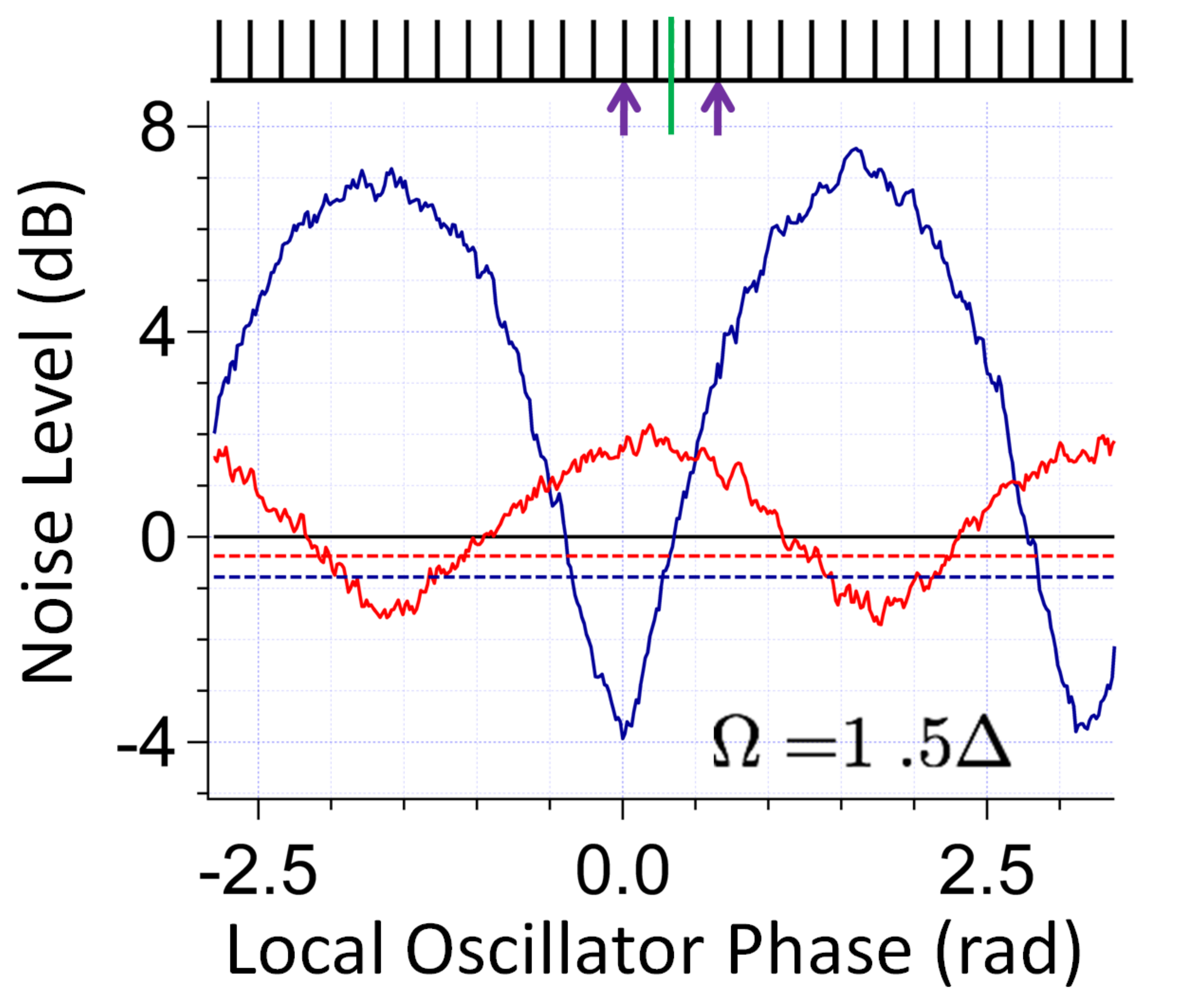}
\caption{Variances $(\Delta A_+)^2$, blue, and $(\Delta A_-)^2$, red, with measurement gains set to 0.3 in lieu of  $1/\Phi=0.618$. (The vLF criterion allows us to deviate from nullifier measurements to find the optimum gain values for maximum violation.) The dashed lines indicate the sub-vacuum-noise violation levels required to prove inseparability.}
\label{fig:vLF}
\end{center}
\vskip -0.2 in
\end{figure}

\paragraph*{Conclusion}
We  demonstrated that the optical frequency comb of a single optical parametric oscillator lives up to its promise as an extremely scalable system for quantum information. We simultaneously generated a record number of quadripartite cluster states, in a record number of Qmodes, all equally entangled. The quantum comb was read by two-tone homodyne detection. Even though the size of the entangled states themselves is not a record, compared to the 14-ion GHZ state \cite{Monz2011}, we demonstrated stringent state preparation requirements for cluster states, a universal quantum computing resource. A practical quantum computer will require an increase in both the number of entangled modes and amount of squeezing. However, the projective measurements required for one-way quantum computing can already be performed on the clusters that we generated \cite{Miwa2010}. Variants of our setup will allow the generation of multiple cube graphs \cite{Zaidi2008}, and a scalable quantum wire and square-grid lattice \cite{Menicucci2008,Flammia2009}.
We thank Nicolas Menicucci, Steven Flammia, Jens Eisert, and G\'eza Giedke for useful discussions. This work was supported by U.S. National Science Foundation grants PHY-0855632 and PHY-0960047. YM was supported by G-COE commissioned by the MEXT of Japan.




\newpage

\onecolumngrid

\markboth{\em\protect{\em Pysher et al.  --- Supplementary Information for ``Parallel generation of quadripartite cluster entanglement in the optical frequency comb''}}{\em\protect{\em Pysher et al.  --- Supplementary Information for ``Parallel generation of quadripartite cluster entanglement in the OFC''}}

\pagestyle{myheadings}

\centerline{\Large Supplementary information for} 
\centerline{\Large``Parallel generation of quadripartite cluster entanglement in the optical frequency comb''}
\begin{center}
Matthew Pysher,$^1$ Yoshichika Miwa,$^{2}$
 Reihaneh Shahrokhshahi,$^{1}$ \\ Russell Bloomer,$^{1}$ and Olivier Pfister$^{1\ast}$\\\ \\
$^1$Department of Physics, University of Virginia, \\ Charlottesville, Virginia 22903, USA\\
$^2$Department of Applied Physics, School of Engineering, The University of Tokyo,\\ 7-3-1 Hongo, Bunkyo-ku, Tokyo 113-8656, Japan\\
$^\ast$opfister@virginia.edu
\end{center}

\date{June 20, 2011}

\maketitle%
\section{Experimental generation of square cluster states in the optical frequency comb}

\subsection{Hamiltonian}
Our optical parametric oscillator (OPO) contained two nonlinear crystals, placed in a two-waist ring cavity.  A 10 mm long periodically poled $\rm KTiOPO_4$ (PPKTP) crystal was placed in the smallest waist.  Three mm of the crystal were poled for quasiphasematching the $zzz$ (``type-0'') parametric downconversion, while the remaining 7 mm were poled for quasiphasematching the $yyz$/$yzy$ (type-II) parametric downconversion, where the first letter denotes the polarization of the annihilated pump photon and the last two letters denote the polarization of the emitted signal (entangled) fields.  The poling periods of 9 $\mu$m for $zzz$ and 458 $\mu$m for $yzy$ were chosen to ensure that both interactions displayed maximum efficiency at exactly the same temperature.  These poling periods were calculated by using the Sellmeier equations given in Ref. \cite{Fan1987}, with the temperature dependences from Ref. \cite{Emanueli2003}.  We operated at $32.34^\circ$C, which was chosen by observing the temperature yielding the largest difference frequency interaction for both $yzy$ and $zzz$.  An unpoled $x$-cut KTP crystal, rotated by $90^\circ$, was placed in the other OPO waist.  This nonphasematched crystal compensated for the birefringence of the PPKTP crystal, thereby making the free spectral range (FSR) of both $y$- and $z$-polarized light an identical $\Delta=945.66$ MHz and allowing for the spectral overlap of these polarization modes.  Modes of both polarizations for each frequency pair were then linked together by parametric downconversion to form square cluster states.    

In the interaction picture, we can write the Hamiltonian of the triply concurrent process taking place in the OPO  
\begin{align}
 H=i\hbar \kappa_{yzy}  \sum_{m}  &\( a_{-n,y}^{\dag} a_{+n,z}^{\dag}+a_{+n,y}^{\dag} a_{-n,z}^{\dag}+\xi\ a_{-n,z}^{\dag} a_{+n,z}^{\dag}\) + H.c. \label{H},
\end{align}
where 
\begin{equation}
\kappa_{yzy}=\frac{2\pi \omega}{n_{y}n_{z}}\chi_{yzy}\beta_{y}
\end{equation}
 and 
\begin{align}
\xi= \frac{\kappa_{zzz}}{\kappa_{yzy}}=\frac{\chi_{zzz}\beta_{z} n_{y}}{\chi_{yzy}\beta_{y} n_{z}}=\frac{d_{yzy} L_{yzy}\beta_{z}n_{y}}{d_{zzz} L_{zzz}\beta_{y}n_{z}}
\end{align}
assuming confocally focused Gaussian beams in the nonlinear crystal \cite{Byer1975}. The optical frequency modes  $( a_{-n,y} , a_{-n,z} )$ and $( a_{+n,y} , a_{+n,z} ) $ are positioned on opposite sides of the comb, with frequencies $\omega_{-n} = \omega_o -(n-1/2) \Delta$ , $ \omega_{n}= \omega_o + (n-1/2) \Delta$ , respectively, where $\Delta= 945.66$ MHz is the free spectral range of the OPO.
 
We can write the Hamiltonian (\ref H) in the general form   
\begin{equation}
H=i\hbar  \kappa_{yzy} \sum_{i,j} G_{ij} \ a_{i}^{\dag} a_{j}^{\dag} + H.c.,
\end{equation}
 where $G$ is of Hankel form and called the adjacency matrix of an $\mathcal H$ (Hamiltonian)- graph
\begin{equation}
G=\left( \begin{array}{cccc}
0 & 0 & 0 & 1 \\
0 & 0 & 1 & 0 \\
0 & 1 & 0 &  \xi \\
1 & 0 & \xi & 0 \end{array} \right).
\end{equation}
Solving the Heisenberg equations for Qmode evolution under ${H}$, we obtain the following squeezed joint operators, or nullifiers:
\begin{align}
& \delta _{\pm} [a_{-n,y}(t) \pm a_{+n,y}^{\dag}(t)]\mp [a_{-n,z}(t) \pm a_{+n,z}^{\dag}(t)] \nonumber\\
& =e^{-r \delta_{\mp}}[\delta _{\pm} (a_{-n,y} \pm a_{+n,y}^{\dag}) \mp (a_{-n,z} \pm a_{+n,z}^{\dag})]
\end{align}
where $ r = \kappa_{yzy}t $ and 
\begin{align}
\delta_{\pm} = \frac{\sqrt{4+\xi^{2}}\pm\xi}{2}.
\end{align}
When $\xi=1$, $\delta_{\pm}=\Phi^{\pm 1}$ ($\Phi$ is the golden ratio) and the real parts of these operators, phaseshifted by $\theta$, coincide with the nullifiers $A_\pm(\theta)$ of the main text.

The experimental setup is displayed in Fig.~2 of the main text. The pump polarization was set to an angle of $19.6^\circ$ with respect to the $y$ axis of the poled crystal, which  meant that 89\% of the light was $y$ polarized and 11\% was $z$ polarized.  This pump polarization orientation was used to counteract the difference in efficiency between the two processes, thereby setting $\xi=1$, and $\delta_{\pm} = \Phi^{\pm 1}$.  It was determined that the SHG efficiency of the $zzz$ interaction in our OPO was 4 times stronger than that of $yzy$ by modeling a scan of resonant second harmonic generation (SHG) output vs infrared (IR) input polarization.  Therefore, in order to match the interaction strengths of the $zzz$, $yzy$, and $yyz$ interactions, it was necessary to pump the cavity with an 8:1 ratio of $y$ light to $z$ light \cite{Boyd}.   

\subsection{Phaselock loop of LO laser}
A small amount of fundamental light at 1064 nm from the Diabolo laser was split off and interfered with light from a Lightwave 126 NPRO 1064 nm laser.  The beat note between these two lasers was detected and electronically mixed with a tunable local oscillator to create an error signal that was subsequently amplified, filtered and sent to a piezo on the lasing crystal of the Lightwave laser, which was capable of tuning the frequency of this laser.  By adjusting the local oscillator frequency, we were able to phase-lock the two IR lasers to any desired frequency difference.  To measure cluster states, we locked the two IR lasers at degeneracy, while sending the remaining portion of the Lightwave beam through a fiber electro-optic modulator (EOM), creating tunable, high-frequency ($\rm< 14 GHz$) sidebands set at $\Omega=\omega_o\pm(n-1/2)\Delta$, where $n$=1,...,15 is an integer.  This modulated field was then sent through another EOM, composed of a $\rm RbTiAsO_4$ (RTA) crystal, which applied FM sidebands at 9.2 MHz .  This doubly modulated beam was split into two, with a small portion being sent to the OPO cavity, where the 9.2 MHz sidebands were used to lock the OPO on the first-order high frequency sidebands with a variant of the Pound-Drever-Hall method \cite{Drever1983}.  The remainder of the beam was sent through a filter cavity of the same FSR as the OPO, which, like the OPO, was locked so as to transmit solely the first-order sidebands from the fiber EOM.  These sidebands were then used as local oscillator (LO) fields in the two-tone balanced homodyne detection (BHD) of each given frequency pair of the OPO. The signal beams emitted by the OPO were aligned with the local oscillators by using an alignment beam which was mode-matched to the OPO through the output coupler.  The reflection of this beam off of the OPO was then interfered with the output from the filter cavity.  The filter cavity's output was aligned to provide maximum homodyne visibility with the alignment beam.  A visibility of 97\% was obtained for the $y$ polarized beam, while 96\% was obtained for the $z$ beam.  The pump beam was then aligned into the OPO in a way that yielded maximum amplification of the resonant modes of the alignment beam.  After the pump beam alignment, the alignment beam was blocked to make way for the OPO locking beam.  

\section{Detection phases of quadrature observables}
\subsection{Nullifier phases}
Our detection system consisted of two balanced homodyne detectors (BHD), one for the $y$ polarized beams and one for the $z$ polarized ones, with all photodiodes measuring both sideband frequencies simultaneously.  These nullifiers were experimentally measured by BHD with the phase-modulated local oscillator beams transmitted through our filter cavity. The fiber EOM modulated the input beam $ e^{ i\omega_ot} $ to output FM beams of the form:
\begin{align}
\alpha_{m} \exp\left\{i\left[ \omega_o +\(n-\frac12\) \Delta\right] t+i\theta_o\right\}-\alpha_m\exp\left\{i\left[ \omega_o -\(n-\frac12\) \Delta\right] t-i\theta_o\right\}
\end{align}
where $\theta_o$ is the phase of the microwave modulation signal at sideband frequency shift $(n-1/2) \Delta$. 
By using a piezo-electric transducer (PZT), we then added a phase shift $\theta_{LO} $  on both sidebands. After reflecting off the PZT mirror, the local oscillator beams became
\begin{align}
 \alpha_{m} \left[ e^{i(  \omega_nt +\theta_o+\theta_{LO})}+e^{i(  \omega_{-n}t -\theta_o+\theta_{LO}+\pi )}\right] = \alpha_{m} \left[e^{i( \omega_nt +\theta_+)}+e^{i(  \omega_{-n}t +\theta_- )}\right]
\label{lobeams}.
\end{align}
From this, we obtain that the detection phases at $\omega_{\pm n}$ were, respectively,
\begin{align}
\theta_+ &= \theta_o + \theta_{LO} \label{theta+}\\
\theta_- &= -\theta_o + \theta_{LO}+\pi.  \label{theta-}
\end{align}
We now consider the variances of the observables measured by our setup
\begin{align}
 A_\pm(g_z,g_y;\theta_-,\theta_+) = & g_y\left[ A_{-n,y}(\theta_-)+ A_{+n,y}(\theta_+)\right] 
 \pm g_z \left[A_{-n,z}(\theta_-)+ A_{+n,z}(\theta_+)\right],
\end{align}
where $g_{y,z}$ are the RF attenuation coefficients applied to the BHD signals of respective polarizations. This is to be compared to the nullifiers that must be measured
\begin{align}
A_\pm(\theta) =  A_{-n,y}(\theta)\mp A_{n,y}(-\theta)\pm\Phi^{\pm1} \left[A_{-n,z}(\theta)\mp A_{n,z}(-\theta)\right].
\label{nulgen1}
\end{align}
It follows that, in order to measure the nullifiers (\ref{nulgen1}), we first needed to realize $\theta_+=-\theta_- + 2p\pi=\theta$, for $A_{-n} (\theta) + A_{n} (-\theta)$, and $\theta_+=-\theta_-+ (2p+1)\pi=\theta$, for $A_{-n} (\theta) - A_{n} (-\theta)$, at each polarization. This leads, respectively, to $\theta_{LO}^{(+)}=(2p-1)\pi/2$ to observe $A_+(\theta)$ and $\theta_{LO}^{(-)}=p\pi$ to observe $A_-(\theta)$. As $\theta_{LO}$ is scanned in the experiment, we therefore expected to observe each nullifier to be squeezed alternatively, every $\pi/2$ rad, which we did, see Fig.~2 in main text. Also, do note that $\theta\neq\theta_{LO}$. Indeed, the aforementioned particular values of $\theta_{LO}$ must be plugged into Eqs.~(\ref{theta+},\ref{theta-}) and the FM phase $\theta_o$ must then be changed by $\pi/2$ in order to yield the noncommuting quadratures---Eqs.~(2-5) in main text. We changed $\theta_o$ by changing the length of the RF cable carrying the FM signal to the EOM.
However, it is essential to note that the nullifier squeezing (\ref{nulgen1}) is independent of $\theta$, and the variation of $\theta_o$ confirmed this by leaving the squeezing signal unchanged at all times.

Secondly, we needed to electronically apply the proper attenuation factors on the two BHD results. The AC output signal of each detector was therefore split into two, so that we could make synchronous measurements of the following: $A_\pm(g_z,0;\theta)$ and $A_\pm(0,g_y;\theta)$. These were useful for van Loock-Furusawa criterion measurements as well. In order to measure nullifiers, we should therefore set $g_z=g_y=\Phi^{-1}=0.618$, or -2.1 dB, in those synchronous measurements. In practice, we had to take into account a 28\% LO loss from the uncoated back surface of a polarizing beam splitter, which affected the $z$ detector only. (The quantum signal was not affected by this loss as it was linearized polarized and incident at the Brewster angle.) We compensated for this LO power difference by adjusting the attenuation factors above by $10\log(1/0.72) = 1.4$ dB.  Thus, instead of attenuating the $z$ channel by -2.1 dB, it was only attenuated by -0.7 dB.  Similarly, the $y$ detector was attenuated by -3.5 dB.     

\subsection{Additional control of nullifiers via polarization}
We introduced an experimental control of the relative phase between the $y$ and $z$ polarized LO beams, by use of a half waveplate (HWP) and a quarter waveplate (QWP) in succession. It is possible to ensure that a subsequent polarizing beam splitter (PBS) will equally split our LO beam by setting the axis of the QWP to $45^\circ$, with respect to the vertical.  The HWP can then be adjusted to give the desired relative phase between the two LO beams.  The application of these waveplates on vertically polarized LO beams, with frequencies $\omega_{n}, \omega_{-n}$  can be explained with Jones calculus.  We can write the effect of the QWP with axis of transmission at $45^\circ$ with the vertical axis as $W_{QWP}$ and the effect of HWP with axis of transmission at $\theta_{HWP}$ with the vertical axis as  $W_{HWP}$ :  
 \begin{align}
W_{HWP} (\theta) = \begin{pmatrix} \cos 2\theta &\sin 2\theta\\ \sin 2\theta& -\cos 2\theta  \end{pmatrix} 
\end{align}
\begin{align}
W_{QWP} (\pi/4) = \begin{pmatrix} 1&-1\\ 1&1 \end{pmatrix} \begin{pmatrix} e^{-i\pi/4}&0\\ 0&e^{i\pi/4} \end{pmatrix}\begin{pmatrix} 1&1\\ -1&1 \end{pmatrix} =\begin{pmatrix} 1&-i\\ -i&1 \end{pmatrix} 
 \end{align}
 Therefore, after effecting Jones matrices on vertical polarized input beam we can write the output beam as:
 \begin{align}
\begin{pmatrix} 1&-i\\ -i&1 \end{pmatrix} \begin{pmatrix} \cos{2\theta} & \sin{2\theta}\\  \sin{2\theta}& -\cos{2\theta} \end{pmatrix}  \begin{pmatrix} 1\\ 0 \end{pmatrix} = -i  \begin{pmatrix} \cos{2\theta}- i \sin{2\theta}\\ -i \cos{2\theta} + \sin{2\theta}, \end{pmatrix} \label{jones}
 \end{align}
where $\theta = \theta_{HWP}$. Eq. (\ref{jones}) shows that for $\theta_{HWP} = \pi/8$,  $\theta_{HWP} = 3\pi/8$, the output beams can be respectively written in matrix form as  $ (1-i)\begin{pmatrix} 1\\ 1 \end{pmatrix} $ and $ - (1+i)\begin{pmatrix} 1\\ -1 \end{pmatrix} $. For these particular values of $\theta_{HWP}$, the local oscillator beam is linearly polarized. Changing $\theta_{HWP}$ by $\pi/4$ can apply a $\pi$ phase shift between the vertical and horizontal polarizations.

\begin{figure}[htbp]
\begin{center}
\includegraphics[width=3in]{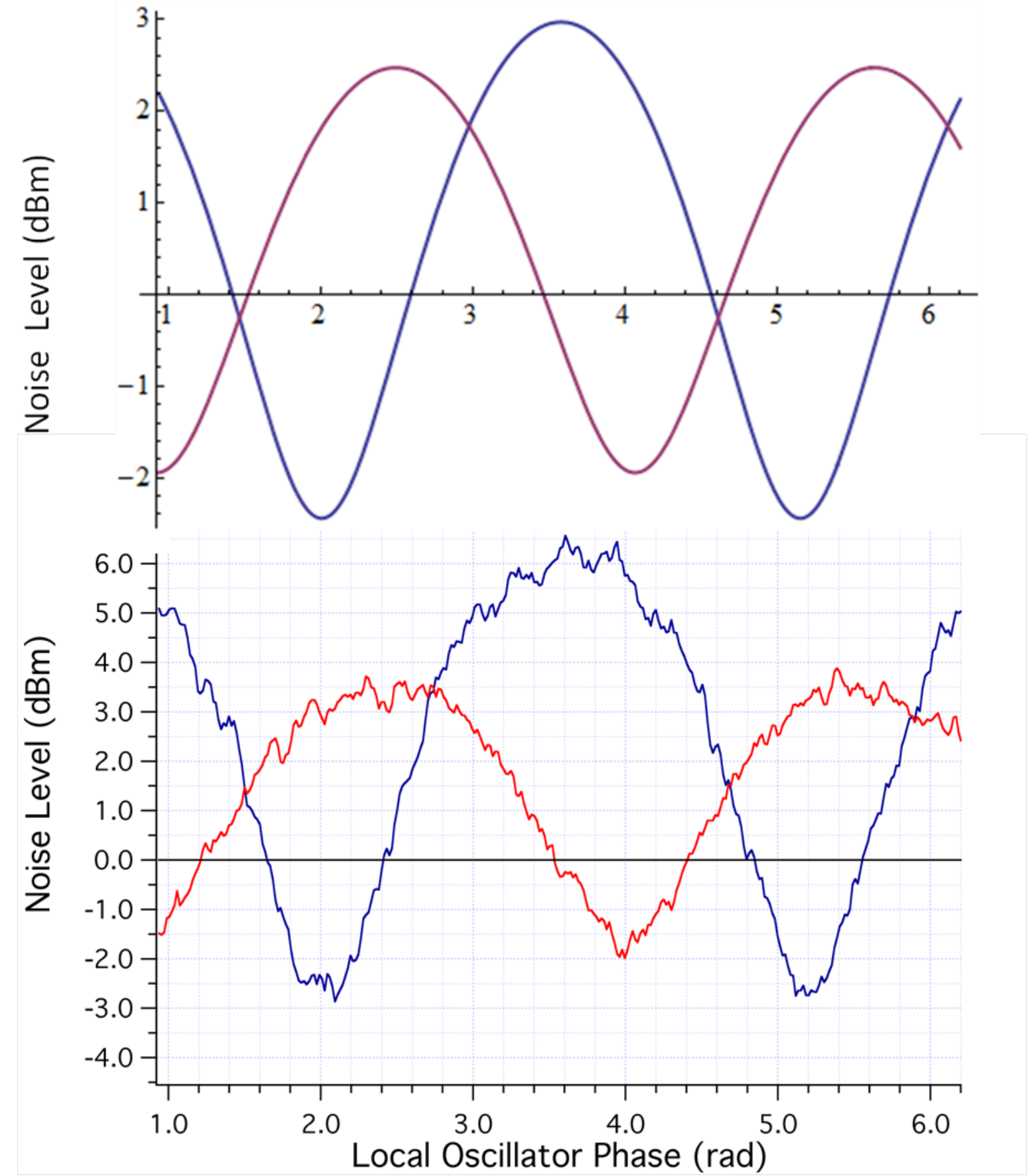}
\caption{Theoretical and experimental variances  for $\theta_{HWP}=\pi/4$. The theoretical trace is for r=0.275.}
\label{fig:nonnull}
\end{center}
\end{figure}

Making simultaneous measurements of the two nullifiers allows us to quickly ensure that $\theta_{HWP}$ is set correctly. As mentioned before, our homodyne measurement yields nullifiers only for  $\theta_{HWP}=\pi/8 + p \pi/2$ If the angle of the HWP is set incorrectly, say with $\theta_{HWP}= \pi/4$, then there is a $\pi/2$ phase shift between the vertical and horizontal polarizations, and the two BHDs will measure quantum amplitudes in quadrature from one another, e.g.\ $A_{-n,y}(\theta)-A_{+n,y}(-\theta)$ and $A_{-n,z}(\theta+\pi/2)-A_{+n,z}(-\theta-\pi/2)$. These expressions can never correspond to nullifiers, but since we are experimentally obtaining finite squeezing, they can,  in some cases, give us squeezing which is comparable to the squeezing of the nullifiers, and it becomes necessary to distinguish such cases from the nullifiers. In \fig{nonnull}, we plot the measured variances in this case. Comparing Fig.~3 in the main text and \fig{nonnull}, it is readily apparent that there is a large difference in the relative phases of the two nullifers for $\theta_{LO}= \pi/8$ vs. $\theta_{LO}= \pi/4$.  
It is important to remark here that, as the amount of squeezing increases, the width of the squeezing hole narrows as the destructive interference that leads to the cancellation of the antisqueezed terms in the nullifier variance becomes more stringent. Therefore, the aforementioned non-optimum cases, already clearly distinguishable in our present conditions, will become all the more obvious.

\section{Crosscorrelations between different cluster states}
In this subsection, we give the calculation result of the variance of nullifiers (\ref{nulgen1}) that involve modes in two different clusters, for example at frequencies $\omega_1$ and $\omega_2$. The results are
\begin{align}
\Delta A_-(\theta) = \left\{\Phi \left[A_{1,y}(\theta)+A_{2,y}(-\theta)\right]- \left[A_{1,z}(\theta)+A_{2,z}(-\theta)\right]\right\}&=\left(1+\frac1\Phi\right) \sqrt{2\cosh \left(\frac{2r}\Phi\right)}
\label{noncor1}\\
\Delta A_+(\theta) = \Delta\left\{\left[A_{1,y}(\theta)-A_{2,y}(-\theta)\right]+ \Phi\left[A_{1,z}(\theta)-A_{2,z}(-\theta)\right]\right\}&=(1+\Phi)\sqrt{2 \cosh (2r\Phi)}
\label{noncor2}
\end{align}
As is clear from Eqs. (\ref{noncor1},\ref{noncor2}), writing nullifiers with the wrong modes can never yield squeezing. Figure 4 in the main text displays the experimental results which confirm these predictions.  We obtained these results by phase-locking the two IR lasers at appropriate frequency differences.  

\section{van Loock-Furusawa inseparability criteria}

In the situation where there is statistical mixing, we can nonetheless prove complete quadripartite nonseparability  by showing that our squeezing levels violate the van Loock-Furusawa (vLF) inequalities \cite{vanLoock2003a}, constructed here from variances of $A_\pm(g_y,g_z;-\theta,\theta)$, where $g_z$ and $g_y$ denote subtraction gains.  In the limit of infinite squeezing, the optimal subtraction gains are $g_z = g_y = 1/\Phi$ as these coincide with the nullifiers. It is important, though, to point out that inseparability is a weaker requirement than state preparation for pure states and the choice of observables above is just an optimal one to ensure inseparability for a mixture; it doesn't constitute a rigorous proof that we have created a cluster state, only a quadripartite entangled state that should nonetheless be very close to a cluster state given that we use cluster-state nullifiers to violate the vLF inequalities. As mentioned previously,  the value of phase $\theta$ Eq.~(\ref{nulgen1}) does not change the squeezing.

In order to derive the sufficient conditions for full inseparability, we may just consider all possible bipartitions of the four modes,
and for each case, we enumerate the necessary conditions for separability.  As the contraposition, we obtain the sufficient condition for full inseparability.

\subsection{One mode - three mode separability}
According to the vLF criteria, if the first mode is separable from the other three modes, the following inequalities are satisfied, 
\begin{align}
\langle Q_-(g_z)^2 \rangle + \langle P_-(g_z)^2 \rangle &\geqslant \frac{1}{2}[|1\times1|+|1\times(-1)+(-g_z )\times(-g_z )+(-g_z )\times g_z |] \notag \\
&= 1, \\
\langle Q_+(g_y)^2 \rangle + \langle P_+(g_y)^2 \rangle &\geqslant g_y^2 , \\
\langle Q_-(g_z)^2 \rangle + \langle P_+(g_y)^2 \rangle &\ (= \langle Q_+(g_y)^2 \rangle + \langle P_-(g_z)^2 \rangle)\ \geqslant g_y, 
\label{eq:calc1}
\end{align}
where $Q_-(g_z)= A_-(0,g_z;0,0)$, etc. The former two of these three equations can be simplified as
\begin{align}
2\langle Q_-(g_z)^2 \rangle &\geqslant 1, \tag{A1} \\
2\langle Q_+(g_y)^2 \rangle &\geqslant g_y^2 \tag{A2},
\end{align}
because of the phase insensitivity of nullifiers, namely $\langle u_i^2 \rangle = \langle v_i^2 \rangle$ with $i=\{ 1,2\}$. 
Equation \eqref{eq:calc1} can be generalized as $\langle Q_-(g_z)^2 \rangle \times a + \langle P_+(g_y)^2 \rangle /a \geqslant g_y$ with any real number $a$,
or equivalently,
\begin{align}
2\sqrt{\langle Q_-(g_z)^2 \rangle \langle Q_+(g_y)^2 \rangle }&\geqslant g_y .\tag{A3}
\end{align}

Similarly, if the third mode is separable from the other three modes, the following are satisfied,
\begin{align}
2\langle Q_-(g_z)^2 \rangle &\geqslant g_z^2, \tag{B1} \\
2\langle Q_+(g_y)^2 \rangle &\geqslant 1, \tag{B2} \\
2\sqrt{\langle Q_-(g_z)^2 \rangle \langle Q_+(g_y)^2 \rangle }&\geqslant g_z .\tag{B3}
\end{align}
In the case that the second (fourth) mode is separable, the same inequalities for separability of the first (third) mode are satisfied.

\subsection{Two mode - two mode separability}
Next, we divide the four modes into two sets of two.
If the first and second modes are separable from the other two modes,
\begin{align}
2\sqrt{\langle Q_-(g_z)^2 \rangle \langle Q_+(g_y)^2 \rangle }&\geqslant g_z + g_y .\tag{C}
\end{align}
Here, we neglect the other two inequalities because they are trivial $(\geqslant 0)$.
In the case that the first and third modes are separable from the other two,
\begin{align}
2\langle Q_-(g_z)^2 \rangle &\geqslant 1+g_z^2, \tag{D1}\\
2\langle Q_+(g_y)^2 \rangle &\geqslant 1+g_y^2, \tag{D2}\\
2\sqrt{\langle Q_-(g_z)^2 \rangle \langle Q_+(g_y)^2 \rangle }&\geqslant |g_z-g_y|.\tag{D3}
\end{align}
In the case that the first and fourth modes are separable from the other two,
\begin{align}
2\langle Q_-(g_z)^2 \rangle &\geqslant 1-g_z^2, \tag{E1}\\
2\langle Q_+(g_y)^2 \rangle &\geqslant 1-g_y^2, \tag{E2}\\
2\sqrt{\langle Q_-(g_z)^2 \rangle \langle Q_+(g_y)^2 \rangle }&\geqslant |g_z-g_y|.\tag{E3}
\end{align}

\subsection{Sufficient conditions for inseparability}
Sufficient conditions for full inseparability can then be obtained by violating one inequality from each group.  The necessary squeezing levels for inseparabilty can be found by normalizing each of the above inequalities by the shot noise level of $\frac{1+g^2}{2}$.  This yields:

\begin{align}
\langle Q_-(g_z)^2 \rangle_N &\leqslant \frac{1}{1+g_z^2}\tag{A1'}, \\
\langle Q_+(g_y)^2 \rangle_N &< \frac{g_y^2}{1+g_y^2} \tag{A2'},\\
\sqrt{\langle Q_-(g_z)^2 \rangle \langle Q_+(g_y)^2 \rangle}_N&< \frac{g_y}{\sqrt{(1+g_z^2)(1+g_y^2)}},\tag{A3'}
\end{align}
\begin{align}
\langle Q_-(g_z)^2 \rangle_N &< \frac{g_z^2}{1+g_z^2}, \tag{B1'} \\
\langle Q_+(g_y)^2 \rangle_N &< \frac{1}{1+g_y^2}, \tag{B2'} \\
\sqrt{\langle Q_-(g_z)^2 \rangle \langle Q_+(g_y)^2 \rangle}_N&< \frac{g_z}{\sqrt{(1+g_z^2)(1+g_y^2)}},\tag{B3'}
\end{align}
\begin{align}
\sqrt{\langle Q_-(g_z)^2 \rangle \langle Q_+(g_y)^2 \rangle}_N&< \frac{g_z+g_y}{\sqrt{(1+g_z^2)(1+g_y^2)}},\tag{C'}
\end{align}
\begin{align}
\langle Q_-(g_z)^2 \rangle_N &< 1, \tag{D1'}\\
\langle Q_+(g_y)^2 \rangle_N &< 1, \tag{D2'}\\
\sqrt{\langle Q_-(g_z)^2 \rangle \langle Q_+(g_y)^2 \rangle}_N&<\frac{|g_z-g_y|}{\sqrt{(1+g_z^2)(1+g_y^2)}},\tag{D3'}
\end{align}
\begin{align}
\langle Q_-(g_z)^2 \rangle_N &< \frac{1-g_z^2}{1+g_z^2}, \tag{E1'}\\
\langle Q_+(g_y)^2 \rangle_N &<\frac{1-g_y^2}{1+g_y^2}, \tag{E2'}\\
\sqrt{\langle Q_-(g_z)^2 \rangle \langle Q_+(g_y)^2 \rangle}_N&<\frac{|g_z-g_y|}{\sqrt{(1+g_z^2)(1+g_y^2)}},\tag{E3'}
\end{align}

For $g_z = g_y = 1/\Phi$, where most of our data was taken, it is necessary that we have
\begin{align}
&\sqrt{\langle Q_-(g_z)^2 \rangle_N \langle Q_+(g_y)^2 \rangle_N }< \frac{2}{\sqrt{5}} \qquad \qquad
\label{eq:fully inseparable conditionsp} 
\end{align}
{and any of the following}
\begin{align}
&\langle Q_-(g_z)^2 \rangle_N < \frac{\sqrt{5}-1}{2\sqrt{5}}, \\
&\langle Q_+(g_y)^2 \rangle_N< \frac{\sqrt{5}-1}{2\sqrt{5}}, \\
&\begin{cases}
\langle Q_-(g_z)^2 \rangle_N &< \frac{1}{\sqrt{5}}, \\
\langle Q_+(g_y)^2 \rangle_N &< \frac{\sqrt{5}+1}{2\sqrt{5}},
\end{cases} \\
&\begin{cases}
\langle Q_-(g_z)^2 \rangle_N &< \frac{\sqrt{5}+1}{2\sqrt{5}}, \\
\langle Q_+(g_y)^2 \rangle_N &< \frac{1}{\sqrt{5}},
\end{cases}
\label{eq:fully inseparable conditionsf}
\end{align}
where the subscript $N$ denotes normalization with the shot noise level ($\frac{1+\delta^2}{2}=\frac{5-\sqrt{5}}{4}$).  Equation ~\eqref{eq:fully inseparable conditionsp} requires that our two nullifers have a mean squeezing value of -0.5 dB, which we easily exceed.  Meanwhile, equation~\eqref{eq:fully inseparable conditionsf} requires -1.4 dB and -3.5 dB squeezing relative to the shot noise level, which we have experimentally achieved for all measured sidebands, thereby ensuring quadripartite entanglement.   
We have enough squeezing to easily satisfy the inequalities given in (B2'), (C'), (D1'), and (D2').  The violations of (A1') and (E2') are, however, much smaller.  It is possible to observe larger violations of the van Loock-Furusawa criteria in these cases by adjusting the gains away from $1/\Phi$.  By setting $g_z=g_y=0.3$, we reduce the necessary squeezing for (A1') from -1.4 dB to -0.4 dB.  Similarly, the necessary squeezing for (E2') is reduced from -3.5 dB to -0.8 dB.  As can be seen in Fig.~6 of the main paper, the squeezing levels at these gains are largely exceeded.

\end{document}